\documentclass[]{article}
\usepackage{amsmath,amssymb}
\usepackage{graphicx}
\makeatletter
\@addtoreset{equation}{section}
\makeatother

\newcommand{\ba}{\begin{eqnarray}}
\newcommand{\ea}{\end{eqnarray}}
\def\half{\frac{1}{2}}
\def\ben{\begin{equation}}
\def\een{\end{equation}}
\def\bea{\begin{eqnarray}}
\def\eea{\end{eqnarray}}
\def\be{\begin{equation}}
\def\ee{\end{equation}}

\def \cM {{\cal M}}\def \cN {{\cal N}}
\def \cL {{\pounds}}
\def\nn{\nonumber}
\def\p{\partial}

\def\ben{\begin{equation}}
\def\een{\end{equation}}
\def\bea{\begin{eqnarray}}
\def\eea{\end{eqnarray}}
\def \nn {\nonumber}

\def \bx {{\bf x}}
\def \p {\partial}

\def\half {\frac{1}{2}} 
\def\ft#1#2{{\textstyle{\frac{\scriptstyle #1}{\scriptstyle #2} } }}
\def\fft#1#2{{\frac{#1}{#2}}}

\def\del{{\partial}}

\newcommand{\hoch}[1]{$\, ^{#1}$}

\newcommand{\auth}{\Large\bf{M. Cveti\v c\hoch{*,**}, G.W. Gibbons\hoch{\dagger,*}
and C.N. Pope\hoch{\ddagger,\dagger}}}

\begin{document}

\begin{flushright}
\hfill {
UPR-1265-T\ \ \ MIFPA-14-29\ \ \
}\\
\end{flushright}

\begin{center}

{\LARGE{\bf Super-Geometrodynamics}}

\vspace{25pt}
\auth

\large

\vspace{30pt}{\hoch{*}\it Department of Physics and Astronomy,\\
University of Pennsylvania, Philadelphia, PA 19104, USA}

\vspace{10pt}{\hoch{\dagger}\it DAMTP, Centre for Mathematical Sciences,\\
 Cambridge University, Wilberforce Road, Cambridge CB3 OWA, UK}

\vspace{10pt}{\hoch{\ddagger}\it George P. \& Cynthia W. Mitchell
Institute for\\ 
Fundamental Physics and Astronomy,\\ Texas A\&M University,
College Station, TX 77843-4242, USA}

\vspace{10pt}{\hoch{**}\it Department of Physics and Astronomy,\\
 Center for Applied Mathematics and Theoretical Physics,
 University of Maribor, Maribor, Slovenia}
\vspace{20pt}

\begin{abstract}

\rm

We present  explicit solutions of the  time-symmetric initial value  
constraints, expressed in terms of freely specifiable  
harmonic functions for  examples of  supergravity  theories, which emerge as effective theories of compactified string theory.  
These  results are a  prerequisite  for the 
study of the time-evolution of  topologically non-trivial 
initial data for supergravity theories,  
thus  generalising the  ``Geometrodynamics''  program of  
Einstein-Maxwell theory to  that of  supergravity theories.
 Specifically, we focus on  examples of  multiple electric Maxwell  and 
scalar fields, and analyse  the initial data problem for the  general
 Einstein-Maxwell-Dilaton  theory both with one and two Maxwell fields, and  the  STU model.  The solutions  are given in terms of 
 up to eight arbitrary harmonic functions in the STU model. 
As a by-product, in order  compare  our  results with known static
solutions, the metric in isotropic coordinates and all the sources of the 
non-extremal black holes are expressed entirely in terms of  
harmonic functions.
We also comment on generalizations to time-nonsymmetric  initial data 
and their relation to cosmological solutions of gauged so-called fake
supergravities with positive cosmological constant.

\end{abstract}

\end{center}

\pagebreak

\tableofcontents

\pagebreak

\section{Introduction}

The consequences  of non-trivial   spacetime  topology 
for   the laws of physics has been  a topic of  perennial interest
for theoretical physicists \cite{Weyl}. 
In its most recent reincarnation 
\cite{Maldacena:2013xja,Maldacena,Susskind:2014jwa}, 
it is the relationship between non-trivial spatial
topology, Einstein-Rosen bridges, wormholes, non-orientable
spacetimes, and quantum-mechanical entanglement
which has been at issue. Not so long ago
 \cite{Thorne:1994xa,Visser:1995cc}, it was the question of
whether such structures would give rise to  
closed timelike curves and the possibility of constructing time machines.    

Such discussions are largely a matter of principle,
since it is unlikely that either astronomical observations
or laboratory  experiments can can shed light on them.
It is important therefore to be sure that the range of such Gedanken 
experiments is restricted by the requirement that they be consistent
with our best current knowledge of the laws of physics.
Thus although  the literature on time travel and wormholes
is replete with models which violate the usual energy
conditions of classical general relativity (cf. \cite{Visser:1995cc}), it is 
more informative to restrict attention  to theories consistent
with this principle, 
and  our current understanding 
of quantum gravity. For these reasons, supergravity theories, in particular those arising  as an effective theories of string and M-theory
are  especially attractive. 
Of course their equations of motion 
include Einstein's vacuum equations 
and the  Einstein-Maxwell equations  as  special cases, and so their 
use does not invalidate  existing work that takes those  into account.
Nevertheless, it is worthwhile to ask what additional 
features arise when  specifically stringy aspects, such as the dilaton
and axion fields, are taken into account. Moreover,  while a great deal is 
now known about supergravity
static and stationary  solutions, such as black holes, 
rather  less is known about time-dependent solutions.

In fact our best information about time-dependent 
wormholes and Einstein-Rosen bridges
comes from a study of the initial value constraints, which
place restrictions on the allowed topology and geometry  of 
possible Cauchy surfaces.   Interestingly,  
the first  hint that Cauchy surfaces in General Relativity 
may be topologically non-trivial came just a year after
the theory's  inception,  with Flamm's \cite{Flamm}
well-known  isometric embedding of the equatorial
plane of the Droste-Schwarzschild solution into Euclidean space
$\mathbb{E}^3$ as the paraboloid of revolution   
\ben
\sqrt{x^2+y^2}= 2M  + \frac{z^2}{8M} \,.
\een
Flamm limited his consideration  to the exterior, $z>0$, 
of what we now call the event horizon, and his illustration
shows only half of the full paraboloid. Einstein and Rosen
\cite{EinsteinRosen} appear  to have  been the first to take seriously
the universe on the other side of what has come to be called
the Einstein-Rosen throat. Later, the study of the time development
of topologically non-trivial initial data was taken up
by Wheeler under the name ``Geometrodynamics'' \cite{Wheeler}. In
in a landmark paper, Misner and Wheeler \cite{MisnerWheeler}
provided examples of simply-connected initial  data 
for both the vacuum Einstein and the Einstein-Maxwell equations, 
with arbitrarily many Einstein-Rosen
throats connecting  many universes to one another.
Misner \cite{Misner:1960zz,Misner2}, followed by Lindquist \cite{Lindquist}, 
 constructed  
non-simply connected examples, called
``wormholes,'' and  Brill and Lindquist \cite{Brill:1963yv} studied 
their energetics. With the development of black hole
theory, it was recognised that the  minimal 2-surfaces 
arising as a consequence of the non-trivial topology provided,
in the time-symmetric case, examples of marginally  trapped surfaces,   
and that these could be used to study the Penrose conjecture
$A\le 16 \pi M^2$  relating the area and mass,  and to 
provide bounds on the amount of gravitational radiation
emitted during the future evolution of the data 
\cite{Gibbons:1972ym,GibbonsSchutz,Cadez,Bishop1,Bishop2}. 

A key notion of the Geometrodynamics programme
was the idea of ``Charge without Charge.'' 
The Maxwell field was taken to be source free, and so a non-vanishing
charge could only arise from ``electric flux lines trapped in the topology
of space.''  With the construction of  ungauged  supergravity theories 
it was realised that the Abelian gauge fields in such theories
were source-free, and so the charges arising therein were therefore  
``central charges'' \cite{Gibbons:1981ux}  and   as a consequence satisfied 
a Bogomol'nyi-Prasad-Sommerfield (BPS) bound  \cite{Gibbons:1982fy}, where the 
 embedding of Einstein-Maxwell theory into $N=2$ supergravity theory was employed.



In this paper we set out to construct  time-symmetric initial date sets  
for  supergravity theories with multiple gauge fields and dilaton-axion fields, focusing on theories in $D=4$ dimensions.  These theories typically arise as a sector of an effective theory of compactified string theories.  A specific ``minimal example'' in this class is the  Einstein-Maxwell-Dilaton  model, with a dilaton-Maxwell coupling constant $a=1$.   While time-symmetric initial data sets with  two arbitrary harmonic functions were constructed by Ortin \cite{Ortin:1995vg},  in this paper we  
 extend and generalise the analysis to  Einstein-Maxwell-Dilaton   models with an arbitrary  
dilaton-Maxwell coupling constant $a$, and  obtain further
initial data sets, now depending on  three arbitrary harmonic functions.   
We compare  these results with those of  known static 
non-extremal black holes,  which we express solely in terms   
harmonic functions.  Furthermore we also generalise these results to the case of Einstein-Maxwell-Dilaton model with {\it two} Maxwell fields. 

An important observation of  Ortin \cite{Ortin:1995vg}, which remains true
for  our solutions,
is that if scalars, such as the dilaton
and hence the string coupling constant, are present, 
they cannot in general be globally defined  
if the the initial manifold is not simply connected, as it would  be
the case for wormhole topologies. This is because 
his explicit solutions for the scalars are not single-valued.
This would seem to have important implications for the considerations
of \cite{Maldacena:2013xja,Maldacena,Susskind:2014jwa}.
This problem may possibly be avoided by considering only
initial data for which the scalars vanish. It would also
not necessarily  be a problem if the scalars were axions.

Our next focus is  on the study of 
time-symmetric initial data  for  the STU supergravity theory, a sector of maximally supersymmetric ungauged supergravity (which is a sector of toroidally compactified string theories), specified by four
 Maxwell fields
$F_I^{\mu \nu}$ (I=1,2,3,4)  and three dilaton-axion fields $a_\alpha+{\rm i} \, e^{-\varphi_\alpha}$ ($\alpha=1,2,3$). Our results are applicable for time-symmetric initial data with four electric fields and three dilation fields turned  on, and   depend on eight arbitrary harmonic 
functions.  In order to compare the initial data problem with the known four-charge electric solutions we express the metric and all the sources  of such black holes in terms of specific harmonic functions. 

The analysis of the  time-symmetric initial data of the  
Einstein-Maxwell-Dilaton models allows us to map the problem to 
that of multi-scalar systems coupled to gravity, which we generalise to 
the case of an arbitrary number $N$ of scalar fields  in Section 5.   
In Section 6  we study the Penrose inequality for the time-symmetric data 
of the  Einstein-Scalar system, and obtain numerical evidence that it 
is always  satisfied.
 We conclude the paper with remarks on interaction energies  for 
time-symmetric initial data. We also comment on generalizations to 
time-dependent data and implications for the study of cosmological  
solutions of gauged supergravities with positive cosmological constant, 
i.e. so-called ``fake supergravities.''

\section{The Initial Value Problem}   

The purpose of this section is to review the formalism for the study of the time-evolution problem for  
theories depending upon a metric $g_{\mu \nu}$, one or more scalars
$\phi_\alpha$, and one or more closed two-forms, or Maxwell fields, 
$F^I = dA^I $ 
whose equations of motion may be obtained from an action functional
$S[g_{\mu \nu}, \phi_\alpha , A^I_\mu]$ that is invariant 
under the semi-direct product of diffeomorphisms and gauge transformations.
For the sake of simplicity of exposition, 
we assume that the Maxwell fields have no sources.
Our  intention here is merely to describe the general framework that
we shall be working with. For more complete  and more rigorous accounts 
the reader is directed to \cite{York,Choquet}.

In Subsection 2.1 we present the evolution equations and derive constraints, and in Subsection 2.2 give the explicit form of constraints. In the Subsection 2.3 we address time-symmetric date and also present the well known explicit results for the vacuum  Einstein gravity and Einstein-Maxwell gravity.
In the subsequent sections  we shall focus on  new results for an  Einstein-Maxwell-Dilaton gravity  model and STU models with multiple scalars and Maxwell fields.

\subsection{Evolution equations and constraints}

Varying the action with respect to $g_{\mu \nu}$ gives
field equations of the form\footnote{We shall use units where $8\pi G=1$.}
\ben
\mathfrak{E}^{\mu \nu } = \sqrt{-g}\, E^{\mu \nu} =
 2 \frac{\,\delta S}{\delta g_{\mu \nu}}   =0\,. \label{metricEOM} 
\een 
Infinitesimal diffeomorphisms
generated by an arbitrary  smooth vector field $V^\mu$ of compact  support
induce a variation of the metric of the form   
\ben
\delta g_{\mu \nu} =  V_{\mu ; \nu}  + V_{\nu ; \mu} \,, 
\een 
where $V^\mu=g^{\mu \nu}V_\nu$, which leaves the action unchanged. 
As a consequence we have the Bianchi-identity
\ben
E^{\mu \nu}{}_{; \nu}= 0 \,. \label{Bianchi}
\een
Similar identities hold for the Maxwell fields: the field
equations take the form   
\ben
{K^I} ^ \mu  = {^*\negthinspace F}^{I \mu \nu}{}_{;\nu} =0  \,,
\qquad {J_I}^\mu =  
{G_I}^{\mu \nu}{}_{;\nu}=0\,, 
\een  
where
\ben
{\mathfrak{G} _I}^{\mu \nu}= \sqrt{-g} \,{G_I}^{\mu \nu}{}_{;\nu}
= - \frac{\, \delta S}{\delta  {F^I}^{\mu \nu }} \,.  
\een
The analogues  of (\ref{Bianchi}) are
\ben
{K^I}^\mu{}_{;\mu}=0 \,, \qquad  {J_I}^\mu{}_{;\mu}=0 \,, 
\label{courrentconseravation}
\een
which may, like (\ref{Bianchi}), be regarded as the consequence of the
invariance of the action under gauge transformations.

    Introducing coordinates $(t,x^i)$, such that the spacetime 
$\{\cM = \mathbb{R} \times \cN ,g_{\mu \nu}\}$ 
is foliated by spacelike hypersurfaces $ \cN$ given by
$t={\rm constant}$, we may write
(\ref{Bianchi}) as 
\ben
\p_t \mathfrak{E}^{\mu t } + \p_i \mathfrak{E}^{\mu i }
+{{\Gamma _i} ^\mu}_j \, \mathfrak{E}^{ij } + 2  {{\Gamma _t}^\mu} _j 
\, \mathfrak{E}^{tj }  +  {{\Gamma _t} ^\mu}_t \, \mathfrak{E}^{tt}
=0\, ,\label{Bianchi2} \een  where ${{\Gamma _\mu}^\sigma}_\nu$
are the Christoffel symbols of the metric $g_{\mu \nu}$.
One sees from (\ref{Bianchi2}) that  the equations (\ref{metricEOM}) 
split into evolution equations
\ben
E^{ij}=0 \label{evolution}
\een
and constraint equations 
\ben
E^{ \mu t} =0\,,\label{constraints} 
\een
such that if the evolution equations (\ref{evolution})
hold for all times, and the constraint equations at some initial time, $t=0$ 
say, then by    (\ref{Bianchi2}) the constraint
equations (\ref{constraints}) will hold for \emph{all} time.    

A similar argument shows that the constraint equations
for the Maxwell fields are given by 
\ben
K^{It} = 0 = J_I^t \,. \label{Gauss} 
\een
The first equation in   (\ref{Gauss}) expresses the absence
of \emph{local} magnetic charge densities and the second, usually called
the Gauss constraint, expresses the absence
of \emph{local} electric  charge densities. The constraint that
$E^{tt}=0$ is usually referred to as the \emph{Hamiltonian constraint}
and the constraint that $E^{ti}=0$ as the \emph{momentum constraint}
or \emph{diffeomorphism constraint}.
For the systems of equations we are considering in this paper
there are no further constraints arising from the scalars $\phi_\alpha$, since
they are not subject to additional gauge invariances.

\subsection{The explicit form of the constraints}

To make progress we need to write out the constraints explicitly
in terms of the metric $g_{ij}$ induced on the initial surface
and some further data including its time derivative $\p_t g_{ij}$.
In our chosen coordinate system $x^\mu=(t, x^i)$, often 
referred to as a \emph{slicing} of spacetime,
the four-dimensional metric takes the form
\ben
g_{\mu \nu} d x ^\mu dx ^\nu = - N^2 dt ^2 + 
g_{ij}(dx^i +N^i dt) (dx^i +N^j dt)
\,. \label{Zermelo} 
\een
All quantities in (\ref{Zermelo}) depend in general on 
all four coordinates. $N$ is a function on $\cN$  called the lapse and 
$N^i$  is a vector field on $\cN$ called the shift.
The coordinates $x^i$ are Lie dragged along the integral curves
or time lines of the  vector field $\frac{\p}{\p t}$. 
The inverse metric is given by
\ben
g^{\mu \nu} \frac {\p}{\p x^\mu  } \otimes  \frac {\p}{\p x^\nu  }
=  \frac{1}{N^2} \bigl 
( \frac{\p}{\p t} - N^k  \frac {\p}{\p x^k } \bigr ) \otimes
( \frac{\p}{\p t} - N^k  \frac {\p}{\p x^k } \bigr )
 + 
g^{ij}  \frac {\p}{\p x^i  } \otimes  \frac {\p}{\p x^j  }\,.
\een 
If the shift vector $N_i$ is non-vanishing,
the vector field $\frac{\p}{\p t}$ is not orthogonal to
the slices $t={\rm constant}$. The unit normal is given by
\ben
n= n^\mu \frac{\p }{\p x^\mu} = 
 \frac{1}{N} \bigl( \frac{\p}{\p t} - N^k  \frac {\p}{\p x^k }   \bigr ) 
\,. \label{normal} 
\een
A full basis for the tangent  bundle 
may be obtained by augmenting $n$ with   an 
orthonormal frame
$e_{\hat i} $ for the Riemannian manifold $\{ \cN, g_{ij} \}$.   
The second fundamental form $K_{ij}$ for the hypersurface
$t={\rm constant}$ is defined by
\ben
K_{ij} = - \half \cL_n g_{ij} \,,  
\een
where $ \cL_n $ denotes the Lie derivative with respect to 
the hypersurface unit vector
field $n$.  For the case of interest to us we have
\ben
E^{\mu \nu}= R^{\mu \nu} - \frac{1}{2} R  g^{\mu \nu} - T^{\mu \nu} 
\,,
\een 
where $T^{\mu \nu}$ is the symmetric energy-momentum tensor
of the matter fields $(\phi_\alpha,F^I_{\mu \nu})$. The Hamiltonian and 
momentum constraints
(\ref{constraints}) thus take the form
\be
 R_{\hat t \hat t }   + \frac{1}{2} R  =  
T_{\mu  \nu } n^\mu n^\nu  \,,\qquad
R_{\mu  \nu} n^\mu e^\mu_{\hat i} = T_{ \mu \nu}   
n^\mu e^\mu_{\hat i}    
\,.\label{constraints2}
 \ee
The left-hand sides of (\ref{constraints2}) may be expressed entirely 
in terms of the
the metric $g_{ij}$, its Ricci scalar $^{(3)}R$ and the second fundamental
form $K_{ij}$ and its covariant derivative $^{(3)}\nabla^k K_{ij}$,
 where $^{(3)}\nabla ^k $ is the  covariant derivative with respect to    
the
metric $g_{ij}$. To do so one uses the Gauss-Coddazi equations,
which relate the Riemann tensor  of $g_{\mu \nu}$
to the Riemann tensor of $g_{ij}$, the second fundamental form
$K_{ij}$ and its covariant derivative.  
One  finally obtains the usual form of the 
Hamiltonian and momentum constraints 
\bea
^{(3)}R  + K^2 - K_{ij}K^{ij} &=& 2 
T_{\mu \nu} \, n^\mu n^\nu  \nn \\
^{(3)} \nabla^j \bigl ( K_{ij} -K  \,g_{ij} \bigr)  &=& 
  T_{ \mu i} \, n^\mu   \,,
\label{constraints3} 
\eea
where 
\be
K= g^{ij}\, K_{ij}\,.
\ee
The initial data for the scalars are simply 
$(\phi_\alpha,\dot \phi _\alpha)$ on the initial time slice, where we define
 $\dot f =  n^\mu \frac{\p f}{\p x^\mu }$ for any function $f$.
Those for the
Maxwell fields are the magnetic  
fields $ {B^I}_{i} = {^*F}^I_{ \mu  i} \,n^\mu $ and 
electric inductions  ${D_I}_{\mu i } = 
-{G_I}_{\mu i } n^\mu $, subject to the constraints (\ref{Gauss}),
which amount to the requirement that both are divergence free,
\ben
^{(3)}\nabla ^i {B^I}_i = 0 = {^{(3)}\nabla}^i {D_I}_i \,. \label{divergence}
\een
 %
 As an example, for the case of the Einstein-Maxwell-Dilaton 
theory an initial data set is  
a seven-tuple 
$\{\cN, g_{ij}, K_{ij}, B_i ,D_i, \phi, \dot \phi \}$  
consisting  of a Riemannian 3-manifold $\{\cN, g_{ij}\} $ and a 
 symmetric tensor field $K_{ij}$,  two functions 
$(\phi ,\dot \phi)$
and two vector fields $B_i$ and $D_i$,  subject to the
(\ref{constraints3}) and (\ref{Gauss}).

\subsection{The time-symmetric case}

A enormous  simplification arises if one assumes that the second fundamental
form of the initial surface, which we take to be at $t=0$, vanishes.
The shift vector $N^i$ also vanishes.
Thus the Hamiltonian and momentum constraints  (\ref{constraints}) reduce to
\be   
^{(3)}R = 2 T_{\hat t \hat t } \,,\qquad 
T_{\hat t \hat i } = 0  \,. \label{tsym}
\ee
In our case, the simplest way to 
arrange that the second equation of (\ref{tsym}) holds is to assume that
\ben
{B^I}_i =0 = \dot \phi_\alpha \,.
\een
The time development of data of this sort will give rise
to a solution which is invariant under $t \rightarrow -t$, and
the spacetime is said to admit a moment of time symmetry.  
>From a dynamical point of view, the system is instantaneously
at rest at $t=0$.

One may now adopt a scheme first proposed by Lichnerowicz 
\cite{Lichnerowicz}.
One assumes that the metric $g_{ij}$ is conformal to some time-independent
background metric ${\bar g} _{ij}$, with  
\ben
g_{ij} = \Phi ^4 {\bar g} _{ij} \,.
\een
The first equation of (\ref{tsym}) now becomes 
\ben
\frac{1}{\Phi ^5} \bigl (  
- 8 {\bar g}^{ij} \, \,
{^{(3)}\bar \nabla}_i \,\, 
{ ^{(3)} {\bar\nabla}  _j}
 + 
{ ^{(3)} {\bar R} }  \bigr )
\Phi  = 2 T_{\mu \nu} n^\mu n^\nu  \,, \label{Lequation}
\een
which is sometimes referred to as  Lichnerowicz's equation.

In principle, Lichnerowicz's method  works for 
any  background manifold  $\{\cN, \bar g_{ij}\}$. In practice
the most useful  cases have been 
\begin{itemize}
\item
The flat metric on Euclidean space $\mathbb{E} ^3$.
This is typically used to give asymptotically flat data.
\item The round metric on the 3-sphere $S^3$. 
This has been used to give initial
data for an inhomogeneous  closed universe.
\item The standard product metric on $S^2 \times S^1$. This has been used
to give initial data for wormholes. 
\end{itemize}

\subsubsection{Vacuum data}

  The simplest case is to set 
\ben
g_{ij} = \Phi^ 4 \delta _{ij} \,, \qquad  \p_i \p_i \Phi =0\,.  
\een
In other words $\Phi$ is a harmonic function on Euclidean space.
We may take
\ben
\Phi = 1 + \sum_{n =1}^N\frac{m_n}{2 |\bx-\bx_n|}\,, \label{1Sdata}
\een
where $\bx=(x_1,x_2,x_3)$ and we assume $m_a>0$. If $N=1$, and 
setting $m_1=M$, 
we obtain the initial data for the 
Schwarzschild solution.
Taking $\bx _1=0$ and writing $\rho=|\bx|$, we can compare with the
Schwarzschild metric in isotropic coordinates, for which $\{\cN,g_{ij}\}$
is manifestly
conformally flat:
\ben    
 ds ^2 = -  \frac{F^2 }{\Phi^2}    dT^2 
+ \Phi ^4  \Bigl \{ d \rho ^2 + \rho ^2 \bigl 
( d \theta ^2 + \sin^2 \theta d \varphi ^2 
 \bigr )    \Bigr \}\,, \label{Sisotropic}  
\een
with 
\ben
F= 1-\frac{M}{2\rho} \,. 
\een
Changing to the  familiar  area coordinate 
\ben
R=\rho \,\Phi^2 \label{areacoordinate} 
\een
transforms the metric (\ref{Sisotropic}) to   
\ben
ds ^2 = -(1- \frac{2M}{R} ) 
\,dT^2 + \frac{dR ^2}{1-\frac{2M}{R} } + 
R^2\bigl(d \theta ^2 + \sin^2 \theta d \varphi ^2 \bigr ) \,.
\een
If we instead take the background 3-metric to be
\ben
\bar g_{ij} dx^i dx^j  = 
 d \chi ^2 + \sin ^2 \chi \bigl( d \theta ^2 + \sin ^2 \theta d \varphi^2 
\bigr )\,  
\een
which is the round metric on $S^3$, and 
solve for a spherically 
solution of (\ref{Lequation}) 
with a simple poles at the north and south
poles of $S^3$, i.e. at $ \chi = 0$ and $\chi=\pi$,
we find
\ben
\Phi  =  \sqrt{M} \sqrt{ \frac{ 1+ \sin \chi}{\sin^2 \chi}}  
=  \frac{\sqrt{M}}{2} \bigl( \frac{1}{\sin \frac{\chi}{2}}  + 
\frac{1}{\cos \frac{\chi}{2}} \bigr ) \,.   
\een
Now setting
\ben
R-M  =  \frac{M}{\sin \chi} = \rho + \frac{M^2}{4\rho} \,, 
\een
one finds that $\Phi^4\,  \bar g_{ij}$ coincides with
the Schwarzschild initial data, i.e. with  (\ref{1Sdata})     
with $N=1$ and $m_1=M$. The event horizon is mapped 
to the equator of $S^3$, i.e. to $\chi=\frac{\pi}{2}$.

Since Lichnerowicz's equation
(\ref{Lequation})  for $\Psi$ is linear in the vacuum case, one may now
superpose solutions, but centred on different points on $S^3$,
as in \cite{Clifton:2014mza}, to obtain initial data 
for a time-symmetric closed universe of black holes 
(c.f.  \cite{LW,Clifton:2013jpa}).  

Finally, if
\ben
\bar g_{ij}dx^i dx^j = 
a^2 \bigl \{  d \mu ^2 +  d \theta ^2 + \sin ^2 \theta
d\varphi ^2   \bigr \}  
\een
and the coordinate $\mu$ is taken to be periodic with period $2\mu_0$,
we obtain the standard product metric on $S^1 \times S^2$.
The function
\ben
\Phi = \frac{1}{\sqrt{\cosh \mu - \cos \theta } } \
\een
satisfies (\ref{Lequation}), and  if
\bea
x &=& a  
\frac{\sin \theta \cos  \varphi}{\sqrt{\cosh \mu - \cos \theta } }\,,\nn \\   
y &=& a 
\frac{\sin \theta \sin \varphi }{\sqrt{\cosh \mu - \cos \theta } }\,,\nn\\ 
z&=& a 
\frac{\sinh\mu  }{\sqrt{\cosh \mu - \cos \theta } } 
\,,
\eea
one finds that  
\ben  
\Phi^4 a^2 \bigl ( d \mu ^2 +  d \theta ^2 + \sin ^2 \theta
d\varphi ^2   \bigr )= dx ^2 + dy ^2 + dz ^2  \,.  
\een
In this case (\ref{Lequation}) is linear \cite{Misner:1960zz}, and we may superpose solutions  as
\ben
\Phi= \sum_{n=-\infty}^\infty \frac{1}{\sqrt{\cosh (\mu + 2 n \mu_0) 
     - \cos \theta  } }\,, 
\een 
 and we obtain Misner's asymptotically-flat wormhole data
on $(S^1\times S^2 ) \setminus \infty$. 
This example may also be obtained using the method of images.
Misner showed also how to obtain  more complicated
non-simply connected examples  using this method \cite{Misner2}.

\subsubsection{Einstein-Maxwell data}
 
Since in this case we have no scalars,   the electric field
$E_i$ is equal to the electric induction $D_i$.
The initial-value constraints therefore 
reduce to
\ben
{ ^{(3)} {\bar R} } = 2 g^{ij} E_iE_j\,, \qquad  ^{(3)}\nabla ^i   E_i =0 
\,.\label{EMconstraints}
\een
For a flat background metric, $\bar g _{ij}=\delta_{ij}$,
Misner and Wheeler \cite{MisnerWheeler} showed that if 
\ben
 \Phi = (CD)^{\half} \,,\qquad   E_i = \frac{D \p_i C-C \p_i D}{CD}
= \p_i\log \frac{C}{D}\,,   
\label{Emsol} 
\een
and $C$ and $D$ are two arbitrary harmonic functions on Euclidean space,
then the initial-value  constraints will be satisfied. Note that
in fact $E_i$ is curl free, 
\ben
\del_i E_j-\del_j E_i =0\,,
\een  
but this is irrelevant as far as the initial-value problem is concerned.
Later we shall see that in more complicated examples
it is not the case that  ${E_I}_i$ is curl-free. 

To obtain regular  initial data for $N$ black holes one chooses
\ben
C= 1 + \sum _{n=1}^N \frac{m_n - q_n }{2 |\bx -\bx_n|} \,,\qquad 
D =  1 + \sum _{n=1}^N \frac{m_n + q_n }{2|\bx -\bx_n| }\,, 
\label{EMdata} 
\een
with $m_n \ge |q_n|$. Taking $N=1$, $\bx_1=0$, $m_1=M$, $q_1=Q$ and 
$|\bx| =\rho$, 
we obtain the initial data for the Reissner-Nordstr\"om metric
in isotropic coordinates,
\ben    
 ds ^2 = -  \frac{E^2F^2}{C^2D^2}  \,  dT^2 
+ C^2 D^2 \Bigl \{ d \rho ^2 + \rho ^2 \bigl 
( d \theta ^2 + \sin^2 \theta d \varphi^2 
 \bigr )    \Bigr \}  \label{RNisotropic}
\een
with 
\bea
C= 1 +\frac{M-Q}{2\rho} \,,&\qquad &D= 1+\frac{M+Q}{2 \rho} 
\nn \\
E= 1 +\frac{\sqrt{M^2-Q^2} }{2\rho} \,,&\qquad& 
F= 1-\frac{\sqrt{M^2-Q^2 } }{2 \rho} \,.
\eea
Using the Schwarzschild area coordinate $R$, given by (\ref{areacoordinate}),
we find the metric takes the standard form
\ben
ds ^2 = -(1- \frac{2M}{R} + \frac{Q^2}{R^2} ) 
\, dT^2 + \frac{dR ^2}{1-\frac{2M}{R} + \frac{Q^2}{R^2} } + 
R^2\bigl(d \theta ^2 + \sin^2 \theta d \varphi ^2 \bigr ) \,.
\een
  
   The initial data for  the  Majumdar-Papapetrou multi-black hole solutions
\ben
ds^2 = - H^{-2} dT^2 + H^2 d \bx ^2 
\een
is obtained by setting $ C= 1$, $D=H$.  To obtain regular
solutions one sets $m_a=q_a$ in (\ref{EMdata}).   

It may be verified that for a  non-flat background metric $\bar g_{ij}$, 
it suffices to replace $C$ and $D$ in (\ref{Emsol}) by 
solutions of 
\ben
\Bigl ( - \, {^{(3)}\bar \nabla} _i \, { ^{(3)} {\bar\nabla} ^i } + 
 \frac{1}{8} \,  { ^{(3)} \bar R}  \Bigr )C =0=
\Bigl ( - \, {^{(3)}\bar \nabla} _i \, { ^{(3)} {\bar\nabla} ^i } + 
 \frac{1}{8} \   {^{(3)}\bar R} \Bigr )D  \,.
\een
For recent work on the numerical evolution
of  Einstein- Maxwell initial data 
the reader is directed to 
\cite{Alcubierre:2009ij,Zilhao:2012gp,Zilhao:2013nda}.

\section{Einstein-Maxwell-Dilaton Theory}\label{emdsec}

  We consider the theory described by the Lagrangian 
\be
{\cal L}= \sqrt{-g}\, (R - 2(\del\phi)^2 - e^{-2a\phi}\, F^2)\,,
\label{emdlag}
\ee
where  coupling $a$ is arbitrary.  The case with $a=1$ is typically considered as a prototype of a sector of an effective theory arising from a  compactification of  string theory. It was this example which  was first addressed for the time-symmetric initial data study in \cite{Ortin:1995vg},  with a  two harmonic function Ansatz. 

We shall  take time-symmetric initial data constraints,  with  the magnetic field
set to zero.  The constants are therefore given by 
\bea
^{(3)} R &=& 2  g^{ij} (\nabla _i \phi  \nabla_j \phi + E_i D_j ) \,,
\label{emdcon1}\\
\nabla^i D_i &=&0\,,\label{emdcon2}
\eea  
where $D_i= e^{-2 a \Phi} E_i$.   

\subsection{Ansatz for initial data using two harmonic functions}

We shall first start with the two-harmonic function  Ansatz. In particular, 
Case (1)   generalizes results  of  \cite{Ortin:1995vg}  to an arbitrary 
coupling $a$.  Case (2) is new, since it allows for electric field which is not a gradient of a potential. 
In the next subsection we will provide a further generalization to three harmonic functions.

The initial data 
Ansatz is
\be
ds_3^2 = \Phi^4\, dx^i dx^i\,,\label{emdmetans}
\ee
with 
\be
\Phi= C^{\frac{1}{4}\gamma}\, D^{\frac{1}{4}\delta}\,,\qquad
e^{-2a\phi}= C^\mu\, D^\nu\label{emdansatz}
\,,
\ee
where $C$ and $D$ will be assumed to be harmonic functions in
the flat 3-metric  $\delta_{ij}$, i.e $\del_i\del_i C=\del_i\del_i D=0$.
The exponents $\gamma$, $\delta$, $\mu$ and $\nu$ will be determined below.
Note that the Ricci scalar $^{(3)}R$ for the Ansatz (\ref{emdmetans}) is
given by
\be
^{(3)}R = -8 \Phi^{-5}\, \del_i\del_i\Phi\,.\label{3R}
\ee
We shall deduce the required form for the initial data for the
electric field by imposing the constraints (\ref{emdcon1}) and
(\ref{emdcon2}).

   Starting with (\ref{emdcon1}), we have
\be
-4\Phi^{-1} \, \del_i\del_i\Phi -(\del_i\phi)(\del_i\phi)=
  e^{-2a\phi}\, E_i E_i\,.\label{Esolve}
\ee
Substituting in the Ansatz (\ref{emdansatz}), where $C$ and $D$ are harmonic,
we seek to write the left-hand side of (\ref{Esolve}) as a perfect square,
\be
\Big(x\, \fft{\del_i C}{C} + y\, \fft{\del_i D}{D}\Big)^2\,,
\ee
which implies the conditions
\be
\gamma(1-\ft14\gamma) - \fft{\mu^2}{4a^2}=x^2\,,\qquad
\delta(1-\ft14\delta) -\fft{\nu^2}{4a^2}=y^2\,,\qquad
\gamma\delta + \fft{\mu\nu}{a^2}= -4 x y\,.\label{perfsq}
\ee
We can then make the natural Ansatz
\be
E_i = e^{a\phi}\, \Big(x\fft{\del_i C}{C} + y\fft{\del_i D}{D}
\Big)\,.
\ee

   The constraint (\ref{emdcon2}), which is $\del_i(\Phi^2\, e^{-2a\phi}\,
   E_i)=0$, implies
\be
\del_i\Big[C^{\ft12\gamma+\ft12\mu}\, 
    D^{\ft12\delta+\ft12\nu}\, 
\Big(x \fft{\del_i C}{C} + y\fft{\del_i D}{D}\Big)\Big]=0\,.
\ee
For harmonic $C$ and $D$, this will be satisfied provided the
terms proportional to $(\del_i C)^2$, $\del_i D)^2$ and 
$\del_i C\, \del_i D$ vanish.  This gives the conditions
\be
(\gamma+\mu-2)\, x=0\,,\qquad (\delta+\nu-2)\, y=0\,,
\qquad (\delta+\nu)\, x + (\gamma+\mu)\, y=0\,,\label{munucond}
\ee
and hence
\be
\mu=2-\gamma\,,\qquad \nu=2-\delta\,,\qquad y=-x\,. 
\ee
The first two equations in (\ref{perfsq}) then imply
\be
(\gamma-\delta)(\gamma+\delta-4) =0\,.
\ee
This has two possible solutions,
\be
\gamma+\delta=4\,,\qquad \hbox{or}\qquad \gamma=\delta\,.
\label{case1case2}
\ee
Let us call these Case (1) and Case (2) respectively. 

\subsubsection{Case (1)}\label{Case1sec}

   For Case (1), where $\gamma+\delta=4$, the third equation in 
(\ref{perfsq}) is automatically consistent with $y=-x$, and hence 
we may write the various exponents in terms of a single parameter
$\alpha$, with
\be
\gamma = 2 -2\alpha\,, \qquad \delta = 2+2\alpha\,,\qquad 
\mu=2\alpha\,,\qquad \nu=-2\alpha\,.
\ee
Thus
we have now satisfied the constraints (\ref{emdcon1}) and (\ref{emdcon2}),
with the initial value data
\be
\Phi^2= C^{1-\alpha}\, D^{1+\alpha}\,,\qquad
e^{-2a\phi}= \Big(\fft{C}{D}\Big)^{2\alpha}\,, 
\qquad
E_i = -\fft{x}{\alpha}\, \del_i\Big(\fft{C}{D}\Big)^{-\alpha}\,,
\label{case1data}
\ee
with
\be
x^2 = 1-\alpha^2 -\fft{\alpha^2}{a^2}\,.
\ee
Note that in this Case (1) example, the electric field $E_i$ can in fact
be written as the gradient of a potential, as in (\ref{case1data}).  This
is not a universal feature, or requirement, for initial value data, as we
shall see in later examples.   The special case when $a=1$ was obtained 
by Ortin \cite{Ortin:1995vg}.  The special case $\alpha=0$ implies
$\phi=0$, and (\ref{case1data}) reduces to the Einstein-Maxwell initial
data discussed in Subsection 2.3.  
The special case where $x=0$ reduces to the
Einstein-Scalar initial data given in \cite{Ortin:1995vg}.

\subsubsection{Case (2)}

  Turning now to the Case (2) example, where $\gamma=\delta$ as in the second
option in (\ref{case1case2}), we may parameterise the indices in terms
of a free constant $\lambda$, with
\be
\gamma=\delta=2\lambda\,,\qquad \mu=\nu=2-2\lambda\,.
\ee
The condition that the third equation in (\ref{perfsq}) be consistent
with $y=-x$ then implies either $\lambda=1$ (in which case the
dilaton vanishes and we are back to the Einstein-Maxwell theory), or else
\be
\lambda = \fft1{1+a^2}\,.
\ee
Thus in Case (2), we have the initial value data
\be
\Phi^2 = (C D)^{\ft{1}{1+a^2}}\,,\qquad
 e^{-2a\phi} = (C D)^{\ft{2a^2}{1+a^2}}\,,\qquad
E_i = \fft1{\sqrt{1+a^2}}\, (C D)^{-\ft{a^2}{1+a^2}}\, 
  \del_i\log\fft{C}{D} \,.
\label{case2data}
\ee
Note that $E_i$ is curl-free only if $a=0$, which reduces to the 
Einstein-Maxwell case.  For all non-zero $a$, the Case (2) initial data
uses an electric field that cannot be written as the gradient of a scalar
potential.  For this reason, it was not obtained in the analysis in
\cite{Ortin:1995vg}.

\subsection{A generalisation with three harmonic 
    functions}\label{3fundatasec}

  Here, we construct an Ansatz for time-symmetric initial data that
depends upon three independent harmonic functions, thus providing a further generalization of the Case (2),  presented in the previous subsection.
   Our motivation for 
seeking this generalisation was provided by considering some known {\it non-extremal}
static black hole solutions   (to be discussed in the next subsection), and
also by considering certain specialisations of the initial data for
STU supergravity (to be discussed in section 4 below).

  Our starting point, with the usual 3-metric $ds_3^2=\Phi^4\, dx^i dx^i$,
is the Ansatz
\be
\Phi= C^{\gamma/4}\, D^{\delta/4}\, W^{\epsilon/4}\,,\qquad
e^{-2a\phi}= C^\mu\, D^\nu\, W^\sigma\,,\label{3fun}
\ee
where the exponents will be determined below. Assuming that $C$, $D$ and 
$W$ are harmonic, we substitute (\ref{3fun}) into 
the left-hand side of (\ref{Esolve}), and seek to write it in the form
\be
\Big(x\, \fft{\del_i C}{C} + y\, \fft{\del_i D}{D} +
    z\, \fft{\del_i W}{W}\Big)^2\,.
\ee
This implies the conditions
\bea
\gamma(1-\ft14\gamma) -\fft{\mu^2}{4 a^2} &=&x^2\,,\qquad
\delta\epsilon + \fft{\nu\sigma}{a^2}= -4 y z\,,\cr
\delta(1-\ft14\delta)-\fft{\nu^2}{4a^2} &=& y^2\,,\qquad
\gamma\epsilon + \fft{\mu\sigma}{a^2} = -4 x z\,,\cr
\epsilon(1-\ft14\epsilon) -\fft{\sigma^2}{4 a^2} &=& z^2\,,\qquad
\gamma\delta + \fft{\mu\nu}{a^2}=-4xy\,.\label{xyzeqns}
\eea
 From (\ref{Esolve}), this leads us to the Ansatz
\be
E_i = e^{a\phi}\, 
\Big(x\, \fft{\del_i C}{C} + y\, \fft{\del_i D}{D} +
    z\, \fft{\del_i W}{W}\Big)
\ee
for the electric field.  

   The constraint (\ref{emdcon2}), which is
$\del_i(\Phi^2\, e^{-2a\phi}\, E_i)=0$, then gives the conditions
\bea
(\gamma+\mu -2)\, x &=&0\,,\qquad (\epsilon+\sigma)\, y + (\delta+\nu)\, z=0\,,
\cr
(\delta+\nu-2)\, y &=& 0\,,\qquad (\epsilon+\sigma)\, x + (\gamma+\mu)\, z=0\,,
\cr
(\epsilon+\sigma-2)\, z &=&0\,,\qquad (\delta+\nu)\, x + (\gamma+\mu)\, y=0\,.
\label{con2eqns}
\eea
It is easy to see that there is no solution where $x$, $y$ and $z$ are all
non-zero.  Without loss of generality, we may therefore proceed by taking
$z=0$.  The equations (\ref{xyzeqns}) and (\ref{con2eqns}) then imply
$y=-x$ and 
\be
\gamma=\delta= \fft{2}{1+a^2}\,,\qquad
\mu=\nu= \fft{2a^2}{1+a^2}\,,\qquad \epsilon=-\sigma= \fft{4a^2}{1+a^2}\,.
\ee
Thus we arrive at the time-symmetric initial data
\bea
\Phi^2 &=& (CD)^{\ft1{1+a^2}}\, W^{\ft{2a^2}{1+a^2}}\,,\qquad
  e^{-2a\phi} = \Big(\fft{CD}{W^2}\Big)^{\ft{2a^2}{1+a^2}}\,,\cr
E_i &=& \fft1{\sqrt{1+a^2}}\, \Big(\fft{CD}{W^2}\Big)^{-\ft{a^2}{1+a^2}}\, 
\del_i \log\fft{C}{D}\,,
\label{3fundata}
\eea
where $C$, $D$ and $W$ are arbitrary harmonic functions. The electric
field is not in general the gradient of a potential function.  The expressions
(\ref{3fundata}) reduce to those of the Case (2) initial data (\ref{case2data})
if the function $W$ is set equal to 1.

\subsection{Some examples of known static solutions}

 Here we examine various examples of known static solutions, and show
how their initial value data fit with the general classes that we 
obtained above.

\subsubsection{Static multi-centre extremal solutions}

   Static multi-centre extremal  solutions in the Einstein-Maxwell-Dilaton
theory were constructed in \cite{Gibbons:1982ih}, and are given by
\bea
ds^2 &=& - C^{-\ft{2}{1+a^2} }dt^2 + C^{\ft{2}{1+a^2}} dx^i dx^i \,,\cr
e^{-2a\phi} &=& C^{\ft{2a^2}{1+a^2} }
\,,\qquad  A_\mu \, dx^\mu= \fft1{C}\, dt\,,
\eea
where $C$ is an arbitrary harmonic function in the flat metric
$dx^i dx^i$.  These solutions are extremal and saturate the BPS bound. 
The electric field in the initial data for this solution is therefore
given by
\be
E_i = - \fft{\sqrt{1+a^2}}{a^2}\, \del_i C^{-\ft{a^2}{1+a^2}}\,.
\ee
Comparing with (\ref{case2data}), we see that the multi-centre metrics
correspond to a specialisation of the Case (2) initial data, in which
the harmonic functions $D=1$.

\subsubsection{Non-extremal static black holes for general $a$}
\label{GibbonsMaedasolutions}

   The theory described by the Lagrangian (\ref{emdlag}) has black hole
solutions given by \cite{gibbmaed}
\bea
ds^2 &=& -\Delta \, dt^2 + \Delta^{-1}\, dr^2 + R^2\, d\Omega_2^2\,,\cr
e^{-2a\phi} &=& F_-^{\ft{2a^2}{(1+a^2)}}\,,\qquad A=q\cos\theta\, d\varphi\,,\cr
\Delta &=& F_+\, F_-^{\ft{(1-a^2)}{(1+a^2)}}\,,\qquad
  R^2 = r^2 \, F_-^{\ft{2 a^2}{(1+a^2)}}\,,\cr
F_\pm &=& 1  - \fft{r_\pm}{r}\,, \qquad q=\sqrt{\fft{r_+\, r_-}{1+a^2}}\,.
\label{gibmaesol}
\eea
We can introduce the isotropic radial coordinate $\rho$, defined by
\be
\log\rho = \int\fft{1}{r\, \sqrt{F_-\, F_+}}\, dr\,,
\ee
which implies that, with a convenient choice for the constant of integration,
\be
r= \rho\, \Big(1+\fft{u^2}{\rho}\Big)\Big(1+\fft{v^2}{\rho}\Big)\,,
\ee
where we have re-parameterised the constants $r_\pm$ as
\be
r_+= (u+v)^2\,,\qquad r_-= (u-v)^2\,.
\ee
In terms of the new quantities, we have
\be
F_- = \fft{\Big(1+\fft{uv}{\rho}\Big)^2}{\Big(1+\fft{u^2}{\rho}\Big)
   \Big(1+\fft{v^2}{\rho}\Big)}\,,\qquad
F_+ = \fft{\Big(1-\fft{uv}{\rho}\Big)^2}{\Big(1+\fft{u^2}{\rho}\Big)
   \Big(1+\fft{v^2}{\rho}\Big)}\,.
\ee
The metric now takes the form
\be
ds^2=-\Delta\, dt^2 + \Phi^4\, (d\rho^2+\rho^2\, d\Omega_2^2)\,,
\ee
where
\be
\Phi^2 = \fft{R}{\rho}= \Big[ \Big(1+\fft{u^2}{\rho}\Big) 
   \Big(1+\fft{v^2}{\rho}\Big)\Big]^{\ft{1}{1+a^2}}\, 
  \Big(1+\fft{uv}{\rho}\Big)^{\ft{2a^2}{1+a^2}}\,,\label{Phiiso}
\ee
and with the dilaton given by
\be
e^{2a\phi}= \Big[ \Big(1+\fft{u^2}{\rho}\Big) 
   \Big(1+\fft{v^2}{\rho}\Big)\Big]^{-\ft{2a^2}{1+a^2}}\, 
  \Big(1+\fft{uv}{\rho}\Big)^{\ft{4a^2}{1+a^2}}\,.\label{diliso}
\ee

   The field strength $F=-q\sin\theta\, d\theta\wedge d\varphi$ has the 
Hodge dual
\be
{*F}= - \fft{q\, \lambda}{f\, \rho^2}\, dt\wedge d\rho\,,
\ee
and hence we can define the dual electric field strength
\be
\widetilde F \equiv e^{-2a\phi}\, {*F} = - \fft{q\sqrt{F_-\, F_+}}{\rho r}\,
  dt\wedge d\rho\,,
\ee
and so
\be
\widetilde F_{t\rho}= -\fft{q}{\rho^2}\, 
   \fft{\Big(1-\fft{uv}{\rho}\Big) \Big(1+\fft{uv}{\rho}\Big)}{
  \Big(1+\fft{u^2}{\rho}\Big)^2\, 
   \Big(1+\fft{v^2}{\rho}\Big)^2}\,.
\ee
The electric field $E_i$ in the initial data can be calculated from
\be
\widetilde F^{\mu\nu}\, \widetilde F_{\mu\nu} = - 2 g^{ij}\, E_i\, E_j\,,
\ee
and hence we find
\be
E_\rho = \fft{ q}{\rho^2}\, \fft{\Big(1+\fft{uv}{\rho}
   \Big)^{\ft{2a^2}{1+a^2}}}{
  \Big[\Big(1+\fft{u^2}{\rho}\Big)\,
   \Big(1+\fft{v^2}{\rho}\Big)\Big]^{\ft{1+2a^2}{1+a^2}}}\,.\label{Eiso}
\ee
Comparing with the general initial data sets that we derived in section
\ref{3fundatasec}, we see that (\ref{Phiiso}), (\ref{diliso}) and 
(\ref{Eiso}) correspond to the special case of (\ref{3fundata}) where
the three harmonic functions are spherically symmetric, and given by
\be
C= 1 + \fft{u^2}{\rho}\,,\qquad D= 1 + \fft{v^2}{\rho}\,,\qquad
   W = 1 + \fft{u v}{\rho}\,.
\ee
(Note that the sign of the dilaton in (\ref{diliso}) is opposite to that
in (\ref{3fundata}).  This is because the spherically-symmetric
solution (\ref{gibmaesol}) 
we are considering here is magnetic rather than electric.  The sign of
the dilaton reverses under dualisation.) 

   It is convenient to re-express the dilaton coupling $a$ in terms of
a parameter $N$, where
\be
a^2 = \fft{4}{N}-1\,.
\ee
The electric field is then given by
\be
E_\rho = \fft{q}{\rho^2}\, \fft{\Big(1+\fft{uv}{\rho}
   \Big)^{2-N/2}}{
  \Big[\Big(1+\fft{u^2}{\rho}\Big)\,
   \Big(1+\fft{v^2}{\rho}\Big)\Big]^{2-N/4}}\,.
\ee
When $N=(1,2,3,4)$ we have $a=(\sqrt3,1,\frac{1}{\sqrt3},0)$, corresponding to
the dilaton couplings when $N$ of the field strengths in the STU supergravity
model are equated, with the remaining $4-N$ set to zero.

   Because these solutions are spherically symmetric, the
electric field in the initial data can always be written in terms 
of a potential, $E_i=-\del_i Z$.
For the $N=2$ and $N=4$ supergravity cases enumerated above, we have
\bea
N=2:\qquad&& Z= -\fft{2q}{(u+v)^2}\, \fft{1-\fft{uv}{\rho}}{
  \Big[\Big(1+\fft{u^2}{\rho}\Big)\,
   \Big(1+\fft{v^2}{\rho}\Big)\Big]^{1/2}}\,,\cr
N=4:\qquad &&  
Z= \fft{q}{u^2-v^2}\, \log\fft{1 +\fft{u^2}{\rho}}{1+\fft{v^2}{\rho}}
\,.
\eea
For general $N$ (integer or non-integer) 
the potentials can also be found in closed form, but
they involve the use of the hypergeometric function:.  
\be
Z= \fft{4q (u+v)^{N/2-3}\, u^{2-N/2}\,\, U^{N/4-1}}{(N-4)(u-v)}
   \,\, _2F_1[\fft{N}{2}-2, 
   \fft{N}{4} -1, \fft{N}{4}; -\fft{v}{u}\,  U]
\,,
\ee
where
\be
U= \fft{1+\fft{u^2}{\rho}}{1+\fft{v^2}{\rho}}\,.
\ee
\subsection{Time-symmetric initial data  with two Maxwell fields}\label{2gaugesec}
We conclude this section by pointing out that one can generalize the time-symmetric initial data results to the case  of Einstein-Maxwell-Dilaton theory with  two Maxwell fields. 
In this case the Lagrangian is of the form:
\be
{\cal L}= \sqrt{-g}\, \Big( R - 2(\del\phi)^2 - e^{-2a\phi}\, F_1^2 
   - e^{-2b\phi}\, F_2^2\Big)\,, \quad b=-\frac{1}{a} \,.\label{2Flag}
\ee
The fixed choice of the dilation coupling $b$ in terms of $a$ is obtained by matching the Lagrangian to a consistent truncation of the STU model with two gauge fields which correspond to   $a=1$, $b=-1$, and $a=\sqrt3$, $b=-\frac{1}{\sqrt3}$, respectively. 

The initial data Ansatz takes the form:
\bea
\Phi^2 &=& (C_1 D_1)^{\ft1{1+a^2}}\, (C_2 D_2)^{\ft{a^2}{1+a^2}}\,,\qquad
e^\phi= \Big(\fft{C_1 D_1}{C_2 D_2}\Big)^{-\ft{a}{1+a^2}}\,,\cr
E^1_i&=& \fft1{\sqrt{1+a^2}}\, 
  \Big(\fft{C_1 D_1}{C_2 D_2}\Big)^{-\ft{a^2}{1+a^2}}\, 
   \del_i\log\Big(\fft{C_1}{D_1}\Big)\,,\cr
E^2_i &=& \fft{a}{\sqrt{1+a^2}}\, 
  \Big(\fft{C_1 D_1}{C_2 D_2}\Big)^{\ft{1}{1+a^2}}\, 
   \del_i\log\Big(\fft{C_2}{D_2}\Big)\,.\label{2Fdata}
\eea
It is interesting to note that if we set the two harmonic functions 
$C_2$ and $D_2$ equal in the above discussion, and, for convenience, define
\be
C_1=C\,,\qquad D_1=D\,,\qquad C_2=D_2=W\,,
\ee
then the initial data given in (\ref{2Fdata}) reduces  precisely to the
initial data (\ref{3fundata}) that we previously derived for the
Einstein-Maxwell-Dilaton system, where $E^1_i=E_i$ and $E^2_i=0$.

These data could in principle also be matched to examples of general static black hole solutions in this  theory.

\section{STU Supergravity}\label{STUsec}

   Four-dimensional STU supergravity is comprised 
of the ${\cal N}=2$ pure supergravity multiplet coupled to three
vector multiplets.  Its gauged version can be obtained as the
consistent truncation of ${\cal N}=8$ gauged $SO(8)$ supergravity to 
its abelian
$U(1)^4$ subsector.  It may also be viewed as ${\cal N}=2$ supergravity
coupled to three vector multiplets.  We shall set the gauge coupling 
constant to zero in our discussion, and focus just on the bosonic 
sector.  There are six scalar fields in total, comprising a dilatonic 
and an axionic scalar in each of the three vector multiplets.  We may 
consistently set the three axionic scalars to zero, provided at the same time
we ensure that their sources, which are proportional to terms of the form
$\epsilon^{\mu\nu\rho\sigma}\, F^I_{\mu\nu}\, F^J_{\rho\sigma}$, are
vanishing.  This can be achieved if we consider field configurations where
the four field strengths have only electric, but not magnetic, components.
 The equations of motion for the remaining fields are then 
described by the Lagrangian\footnote{We are using the customary 
normalisations 
for the kinetic terms of STU supergravity here, which are smaller by a factor
of 4 than those we have used for the other theories discussed in this paper.}
\be
{\cal L} = \sqrt{-g}\, \Big[ R - 
\ft12 \sum_{\alpha=1}^3 (\del\varphi _\alpha )^2 -
  \ft14 \sum_{I=1}^4 X_I^{-2}\, (F^I)^2 \Big]\,,\label{STUlag}
\ee
\be
X_I= e^{-\ft12 {\bf a}_I\cdot{\boldsymbol \varphi}}\,,\qquad
{\boldsymbol \varphi}= (\varphi_1,\varphi_2,\varphi_3)\,,
\label{Xdef}
\ee
and we define
\be
{\bf a}_1=(1,1,1)\,,\qquad {\bf a}_2=(1,-1,-1)\,,\qquad 
{\bf a}_3=(-1,1,-1)\,,\qquad {\bf a}_4=(-1,-1,1)\,.\label{adef}
\ee

   The constraints for time-symmetric initial data will then be
given by
\bea
  {^{(3)}R}&=& \ft12 g^{ij}\, \sum_\alpha \del_i\varphi_\alpha 
\del_j\varphi_\alpha 
   + \ft12 g^{ij}\,  \sum_I X_I^{-2}\, E^I_i E^I_j\,,\label{STUcon1}\\
 {^{(3)}\nabla}^i(X_I^{-2}\, E^I_i)&=&0\,.\label{STUcon2}
\eea

\subsection{Time-symmetric initial data}\label{STUidsec} 

We start with $8$ arbitrary harmonic functions $C_I,D_I$, $I=1,2,3,4$
on Euclidean space $\mathbb{E} ^3$ .
The 3-metric is assumed to be given by 
\ben
ds^2 = \Pi^{\frac{1}{2}} d {\bf x}^2  \label{eqn3}
\een
where we have defined
\be
\Pi\equiv \prod_{I=1}^4  C_ID_I\,.
\ee
Thus we have
\be
\Phi^2= \Pi^{1/4}\,.\label{PhiSTU}
\ee
The scalars are given by 
\ben
X_I= \frac{\Pi^{1/4}}{C_I D_I}\,.\label{XIs}
\een
In this example, we cannot in general
express the electric fields $E^I_i$ in terms of
scalar potentials.  A simple way to obtain expressions for $E^I_i$
that are consistent with the constraints is first to substitute 
(\ref{PhiSTU}) and (\ref{XIs}) into the constraint (\ref{STUcon1}), since this
leads us to a natural conjecture for $E_i^I$.  Noting from (\ref{Xdef}) and
(\ref{adef}) that
\be
\del_i\varphi_\alpha \del_i\varphi_\alpha  = 
\sum_{I=1}^4 \Big(\fft{\del_i X_I \, \del_i X_I}{X_I^2}\Big)\,,
\ee
we find after a little algebra that
\be
\Phi^4\, \Big(\, ^{(3)}R -\ft12 g^{ij}\, 
\del_i \varphi _\alpha \del_j\varphi_\alpha \Big) =
  \ft12 \sum_{I=1}^4 \Big(\fft{\del_i C_I}{C_I} -\fft{\del D_I}{D_I}\Big)^2\,.
\ee
Thus the constraint (\ref{STUcon1}) is satisfied if we take
\be
E_i^I = X_I\, \del_i \log\Big(\fft{C_I}{D_I}\Big)=
  \fft{\Pi^{1/4}}{C_I D_I}\, \del_i \log\Big(\fft{C_I}{D_I}\Big)\,.
\label{ESTU}
\ee
It remains to verify that the constraints (\ref{STUcon2}) are satisfied.
Thus we have
\be
\del_i(\Phi^2\, X_I^{-2}\, E^I_i)= \del_i\Big[C_I D_I\, 
\del_i \log\Big(\fft{C_I}{D_I}\Big)\Big]= D_I \nabla^2 C_I -
 C_I \nabla^2 D_I\,,
\ee
which indeed vanishes because $C_I$ and $D_I$ are harmonic. 
It is easy to verify that the curls of the electric fields $E_i^I$ are
non-vanishing, and so it is not possible to write them as the gradients
of any potentials.

  There are four special cases of the STU supergravity initial data
that reduce to data for the Einstein-Maxwell-Dilaton system discussed
in section \ref{emdsec}; in particular, they all fit into the 
initial data with three harmonic functions, which we derived in section
\ref{3fundatasec}.  They correspond to the truncations of the
STU theory to the Einstein-Maxwell-Dilaton theory with $a=\sqrt3$, 1,
$\frac{1}{\sqrt3}$ and 0.  Modulo permutation choices, the specialisations
of the initial data are:
\bea
a=\sqrt3:&&\quad C_1=C\,,\qquad D_1=D\,,\qquad 
  C_2=C_3=C_4=D_2=D_3=D_4=W\,,\cr
a=1:&&\quad C_1=C_2=C\,,\qquad D_1=D_2=D\,,\qquad C_3=C_4=D_3=D_4=W\,,\cr
a=\fft1{\sqrt3}:&&\quad C_1=C_2=C_3=C\,,\qquad
   D_1=D_2=D_3=D\,,\qquad C_4=D_4=W\,,\cr
a=0:&&\quad C_1=C_2=C_3=C_4=C\,,\qquad D_1=D_2=D_3=D_4=D\,,\qquad W=1\,.
\eea
 There are two consistent truncations (modulo permutation)  of the  STU model  to  the Einstein-Maxwell-Dilaton theory with two  Maxwell fields  (\ref{2Flag}) and  the following dilation couplings:
 $a=1$, $b=-1$ and $a= \sqrt3$, $b=-\frac{1}{\sqrt3}$. These truncations result in four independent
harmonic functions
$C_1$, $D_1$, $C_2$ and $D_2$ remaining, namely
\bea
a=1, \ b=-1:&&\qquad 
  C_1=C_3\,,\qquad\qquad D_1=D_3\,,\qquad C_2=C_4\,,\qquad D_2=D_4\,,\cr
 a= \sqrt3, \ b=-\frac{1}{\sqrt3}:&&\qquad C_2=C_3=C_4\,,\qquad D_2=D_3=D_4\,.
\eea

\subsection{Examples of known static solutions}

Here we look at various examples of known static solutions in the
STU supergravity theory, and show how their initial value data correspond to
special cases of the above 8 harmonic function initial data.  As a by-product we express the  non-extremal static black hole spatial metric and all the sources in terms of   eight specific harmonic functions.

\subsubsection{Extremal multi-centre black holes}\label{multiextremal}

   The general static extremal multi-centre black holes are given by
\bea
ds^2 &=& -\Big(\prod_{I=1}^4 C_I\Big)^{-1/2}\, dt^2 + 
  \Big(\prod_{I=1}^4 C_I\Big)^{1/2}\, dx^i dx^i\,,\cr
X_I &=& \Big(\prod_{I=1}^4 C_I\Big)^{1/4}\, C_I^{-1}\,,\qquad 
    A^I_\mu dx^\mu = -C_I^{-1}\, dt\,,
\eea
where the $C_I$ are arbitrary harmonic functions in the flat transverse
metric $dx^i dx^i$.  From this, we see that the electric fields $E^I_i$ in
the initial data are given by
\be
E_i^I= \Big(\prod_{I=1}^4 C_I\Big)^{1/4}\, C_I^{-2}\, \del_i C_I\,.
\ee
These solutions saturate the BPS bound.  The single centered solution was first obtained \cite{CY} in $N=4$ supergravity and preserves $\frac{1}{4}$ of supersymmetry.

Comparing these solutions with the initial data (\ref{PhiSTU}), (\ref{XIs})
and (\ref{ESTU}), we see that they correspond to the case
$D_I=1$. 

\subsubsection{Static spherically-symmetric non-extremal black holes}

The static spherically symmetric solutions, first given in \cite{CYII},  take the form:
\bea
ds^2 &=& - \bigl(\prod_{I=1}^4  H_I \bigr )^{-\half}  
(1- \frac{2m}{r} ) dt^2 +  
\bigl(\prod_{I=1}^4  H_I \bigr )^{\half}
\Bigl \{  \frac{dr^2}{ 1- \frac{2m}{r}    }   
+r^2 d\Omega^2 \Bigr \}\,,\cr \label{slikecoords}
H_I &=& 1+ \frac{2m \sinh^2 \delta_I} {r} \, \  
X_I = \Big(\prod_{I=1}^4 H_I\Big)^{1/4}\, H_I^{-1}\,,\ \ A^I_\mu dx^\mu = (1-H_I^{-1})\, \coth\delta_I\, dt\,.\label{4chbh}
\eea
We define the isotropic radial  coordinate $\rho$ by
$r=\rho+m+ \ft{m^2}{4 \rho} $ and find that 
\ben
\frac{dr^2}{1- \frac{2m}{r}}   +
r^2 \bigl( d \theta ^2 + \sin ^2 \theta d \phi ^2 \bigr )   = \Bigl(1 + \frac{m}{2\rho} \Bigr ) ^4
\Bigl \{  d \rho ^2 + \rho ^2 d\Omega^2   \Bigr \} \label{isocoords}
\een
We now find that 
\ben
(1+ \frac{m}{2\rho})^2   H_I=C_ID_I\,, 
\een
where $C_I$ and $D_I$ are spherically symmetric harmonic functions.
\footnote{Note that unless $m=0$, $H_I$ are not harmonic functions
with respect either of the metrics in braces in (\ref{slikecoords}) or
(\ref{isocoords}). In this paper we have used $C$ and $D$,
possibly subscripted, to denote generic harmonic functions.
}  
\ben
C_I= 1 + \frac{me^{2 \delta _I} }{2\rho} \,,\qquad 
D_I= 1 + \frac{me^{-2 \delta _I}} {2\rho} \,.\label{CIDIsphsym}
\een
Note that $C_I$ and $D_I$, unlike $H_I$ itself, are harmonic in the
flat transverse 3-metric $d\rho^2+ \rho^2 d\Omega^2$.

   In terms of the isotropic radial coordinate, the metric (\ref{4chbh})
becomes
\be
ds^2 = -\Pi^{-1/2}\, f_+^2\, f_-^2\, dt^2 + \Pi^{1/2}\, (d\rho^2 +
      \rho^2 d\Omega^2)\,, \label{ISOTROPICUNGAUGED}
\ee
where we have defined
\ben
\Pi=  \prod_{1\le I\le 4} C_I  D_I\,,\qquad f_\pm= 1 \pm \fft{m}{2\rho}\,.
\een
The scalar fields and gauge potentials can be written as
\be
X_I = \fft{\Pi^{1/4}}{C_I D_I}\,,\qquad A^I_\mu dx^\mu =
   \Big(-\fft{1}{C_I} + \fft{1}{D_I}\Big)\, dt\,.
\ee
The electric fields $E^I_i$ on the initial data surface
$t={\rm constant}$  are purely radial, and may be obtained by noting that
\ben
g^{ij } E^I_i  E^I_j = -\half F^I_{\mu \nu}  {F^I}^{\mu \nu} \,.
\een
The result is that
\ben
E^I_\rho= -\frac{\Pi^{\frac{1}{4}}  2m \sinh \delta_I \cosh \delta_I}
{\rho^2 C_I^2 D_i^2 } \,.\label{Erho}
\een

   It is now straightforward to verify that the initial data for this
solution, given by $\Phi^2=\Pi^{1/4}$ and $X_I=\Pi^{1/4}/(C_I D_I)$, and
with $E^I_i$ given by (\ref{Erho}), corresponds to the special case
of the general 8-function initial data (\ref{PhiSTU}), (\ref{XIs}) and
(\ref{ESTU}) where $C_I$ and $D_I$ are spherically symmetric and given by
(\ref{CIDIsphsym}).

\section{Multi-Scalar Systems Coupled to Gravity}

  In this section we present some additional examples of time-symmetric
initial for systems of scalar fields coupled to gravity.  We begin
by showing how all the cases we have discussed so far, involving one or
more Maxwell fields, can be mapped into systems describing Einstein gravity
coupled purely to scalar fields.  The essential feature that allows this 
mapping is that in all the previous examples, the electric fields in the
initial data are either expressible as the gradients of scalar functions,
or else they are are proportional to the gradients of scalar functions.
We also give a direct construction of a general new class of examples 
of Einstein gravity coupled to a system of scalar fields, and we show how
the Einstein-Scalar systems obtained as mappings from systems with
Maxwell fields are all special cases within this broader class. 

\subsection{Mapping Maxwell data to scalar data}\label{maxwelltoscalars}

     It has been observed previously that the time-symmetric 
initial value problem
for the Einstein-Maxwell system can be mapped into an equivalent 
initial value problem involving only scalar fields \cite{Ortin:1995vg}.  Let us
consider for simplicity the case where the magnetic field vanishes, and
hence the initial value constraints are given by  (\ref{EMconstraints}). 
Since $E_i$ is the gradient of a scalar in this case we can define
$E_i=\del_i\psi$, and so the Ricci constraint in (\ref{EMconstraints})
becomes
\be
^{(3)}R= 2 g^{ij}\, \del_i\psi \del_j\psi\,.
\ee
This constraint is solved by the writing $\Phi$ and the scalar 
field $\psi$ in terms of the two harmonic functions $C$ and $D$
as
\be
\Phi^2= CD\,,\qquad \psi = \log\fft{C}{D}\,.
\ee

  We may now observe that a similar mapping of Maxwell data into data
for a scalar field may be made in the more complicated theories that we
have considered in this paper. For the Einstein-Maxwell-Dilaton system 
discussed in section \ref{emdsec}, the Ricci constraint 
(\ref{emdcon1}) for the Case (1) data in equation (\ref{case1data}) may
be rewritten as
\be
^{(3)}R = 2g^{ij}(\del_i\phi\del_j\phi + \del_i\psi\del_j\psi)\,,
\label{2scalcon}
\ee
where we have written 
\be
E_i= \Big(\fft{C}{D}\Big)^{-\alpha}\, \del_i\psi
\ee
and the fields $\Phi$, $\phi$ and $\psi$ are expressed in terms of the
harmonic functions $C$ and $D$ as
\be
\Phi^2= C^{1-\alpha}\, D^{1+\alpha}\,,\qquad 
   \phi=-\fft{\alpha}{a}\log\fft{C}{D}\,,\qquad \psi= x\, \log\fft{C}{D}\,,
\label{2hid}
\ee
where $x=\sqrt{1-\alpha^2 -\alpha^2/a^2}$. 
This provides initial data for the theory of Einstein gravity coupled to
two scalar fields, described by the Lagrangian
\be
{\cal L}_4 = \sqrt{-g}\, \Big(R -2 (\del\phi)^2 -2(\del\psi)^2\Big)\,.
\label{2scallag}
\ee
The constants $\alpha$ and $a$ in (\ref{2hid}) are arbitrary parameters
that may be chosen when specifying the initial data.

  For the solution of the initial data for the Einstein-Maxwell-Dilaton system
using three harmonic functions $C$, $D$ and $W$, discussed in section 
\ref{3fundatasec}, we may write
\be
E_i =(CD)^{-\ft{a^2}{1+a^2}}\, W^{\ft{2a^2}{1+a^2}}\, \del_i\psi \,,
\ee
and reinterpret the initial value problem as again being that for 
 Einstein gravity coupled to two scalar fields, described by (\ref{2scallag}),
with the initial constraint
(\ref{2scalcon}), and satisfied by the initial data
\be
\Phi^2= (CD)^{\ft1{1+a^2}}\, W^{\ft{2a^2}{1+a^2}}\,,\qquad
  \phi=-\fft{a}{1+a^2}\, \log\fft{CD}{W^2}\,,\qquad
  \psi=\fft1{\sqrt{1+a^2}}\, \log\fft{C}{D}\,.
\ee
 
   Note that in both of the above examples the electric field
$E_i$ in the initial value data for the original Einstein-Maxwell-Dilaton
theory is not curl-free, and thus cannot itself be written as the gradient
of a scalar.  Nonetheless, $E_i$ is in each case proportional to a gradient,
and that enables us map the initial value problem into one with a second 
scalar field instead of the electric field.

   The initial value data that we obtained in section \ref{STUsec} for
time-symmetric solutions of STU supergravity can also be mapped into
data for an Einstein-Scalar system, this time with a total of seven 
scalar fields.  We do this by noting from (\ref{ESTU}) that we may
write
\be
E^I_i= \fft{\Pi^{1/4}}{C_I D_I}\, \del_i\psi_I\,,
\ee
for which the initial value Ricci constraint (\ref{STUcon1}) becomes
\be
^{(3)}R = \ft12 g^{ij}(\del_i\varphi_\alpha \del_j\varphi_\alpha +
   \del_i\psi_I \del_j\psi_I)\,,
\ee
with the initial data being given by
\be
\Phi^2 =\Pi^{1/4}\,,\qquad X_I= \fft{\Pi^{1/4}}{C_I D_I}\,,\qquad
  \psi_I = \log\fft{C_I}{D_I}\,,
\ee
where $\Phi=\prod_I (C_I D_I)$.
Thus we have initial data in the form of eight arbitrary harmonic functions 
$(C_I,D_I)$ for the system of seven scalar fields $(\varphi_\alpha,\psi_I)$
coupled to gravity, and described by the Lagrangian
\be
{\cal L}_4=\sqrt{-g}\Big( R -\ft12(\del\varphi_\alpha)^2 -\ft12(\del\psi_I)^2
\Big)\,.
\ee

   Finally, we may consider the theory of Einstein-Scalar gravity coupled 
to two gauge fields, which was described in section \ref{2gaugesec}.  We 
showed that the theory described by the Lagrangian (\ref{2Flag}), with 
$b=-1/a$, admits the time-symmetric initial data given in (\ref{2Fdata}).
We may therefore introduce two scalar fields $\psi_1$ and $\psi_2$, 
such that
\bea
E_i^1&=& \Big(\fft{C_1 D_1}{C_2 D_2}\Big)^{-\ft{a^2}{1+a^2}}\, \del_i\psi_1\,,
\cr
E_i^2&=&  \Big(\fft{C_1 D_1}{C_2 D_2}\Big)^{\ft{1}{1+a^2}}\, \del_i\psi_2\,,
\eea
and thus obtain the time-symmetric initial data
\bea
\Phi^2 &=& (C_1 D_1)^{\ft1{1+a^2}}\, (C_2 D_2)^{\ft{a^2}{1+a^2}}\,,\qquad
\phi= -\fft{a}{1+a^2}\, \log\fft{C_1 D_1}{C_2 D_2}\,,\cr
\psi_1 &=& \fft1{\sqrt{1+a^2}}\, \log\fft{C_1}{D_1}\,,\qquad
\psi_2= \fft{a}{\sqrt{1+a^2}}\, \log\fft{C_2}{D_2}\,,
\eea
using the four harmonic functions $C_1$, $D_1$, $C_2$ and $D_2$, 
for the theory of three scalar fields coupled to gravity, described by the
Lagrangian
\be
{\cal L}_4 = \sqrt{-g}\, \Big(R - 2(\del\phi)^2 - 2(\del\psi_1)^2 - 
          2 (\del\psi_2)^2\Big)\,.
\ee

\subsection{Einstein gravity coupled to $N$ scalar fields}

   Here we present a direct construction of time-symmetric initial data
for a system of $N$ scalar fields coupled to gravity, and described by the
Lagrangian
\be
{\cal L}= \sqrt{-g}\, (R -2 \del\phi_I \del\phi_I)\,,
\ee
where $1\le I \le N$. The time-symmetric initial value constraint is then
\be
  ^{(3)}R = 2 g^{ij} \del_i\phi_I \del_j\phi_I\,.\label{ivcon}
\ee
We make the Ansatz 
\be
\Phi= \prod_{a=1}^M C_a^{n_a}\,,\qquad 
        \phi_I = 2\sum_{a=1}^M m_{aI}\, \log C_a\,,\label{scalarans}
\ee
where $C_a$ for $1\le a\le M$ are $M$ harmonic functions.

  Plugging into (\ref{ivcon}) implies the following constraints on the
constants $n_a$ and $m_{aI}$:
\bea
a\ne b:&& n_a n_b + m_{aI} m_{bI}=0\,,\\
a=b:&& n_a(n_a-1) + m_{aI} m_{aI}=0\,.\label{solcon}
\eea
(Summation over $I$ is understood in each case.)  This implies a total of
$\ft12M(M+1)$ constraints on the total of $M(N+1)$ constants.  
If we define the $(N+1)$-component vectors
\be
{\bf m}_a=(m_{a1},m_{a2},\ldots, m_{a N}, n_a)\,,
\ee
then the conditions in (\ref{solcon}) can be written as
\be
{\bf m}_a\cdot {\bf m}_b= n_a\, \delta_{ab}\qquad (\hbox{no sum on}\ a)\,.
\ee
Defining ${\bf Q}_a={\bf m}_a/\sqrt{n_a}$, we then have
\be
{\bf Q}_a\cdot {\bf Q}_b=\delta_{ab}\,.\label{QQ}
\ee
Thus we obtain a solution for every choice of orthonormal $M$-frame in 
${\mathbb R}^{N+1}$.  We must therefore have $M\le N+1$.

   Considering the maximal case $M=N+1$, $SO(N+1)$ acts on the 
orthonormal bases for ${\mathbb R}^{N+1}$, but the $SO(N)$ subgroup acting
on the first $N$ components merely rotates the $N$ scalar fields into 
themselves, and produces a physically indistinguishable solution.  If one
acts with an element of $SO(N+1)$ that is not contained in $SO(N)$, the
scalar fields $\phi_I$ will mix with the scalar $\sigma=2\log\Phi$ and hence
give rise to a geometrically distinct solution of the initial-value
constraint (\ref{ivcon}).  The
space of inequivalent solutions is therefore given by the coset
$SO(N+1)/SO(N)$, which is isomorphic to $S^N$.   

  The various Einstein-Scalar theories we obtained in section 
\ref{maxwelltoscalars}  are all
examples encompassed within the above discussion.  We have

\begin{itemize}

\item[$\bullet$] Einstein-Dilaton: $(M,N)= (2,1)$

\item[$\bullet$] Einstein-Maxwell-Dilaton: $(M,N)= (3,2)$

\item[$\bullet$] STU Supergravity:  $(M,N)=(8,7)$

\item[$\bullet$] Einstein-Dilaton+2 Maxwell: $(M,N)=(4,3)$

\end{itemize}

  For example, in the case of STU supergravity, one finds, after defining 
\be
\phi_I=\ft12 (\psi_1,\psi_2,\psi_3,\psi_4,\varphi_1,\varphi_2,\varphi_3)\,,
\qquad C_a=D_{a-4} \ \hbox{for}\ 5\le a\le 8\,,
\ee
that 
\bea
{\bf m}_1&=&\ft12(1,0,0,0,\ft12,\ft12,\ft12,\ft12)\,,\qquad
{\bf m}_2=\ft12(0,1,0,0,\ft12,-\ft12,-\ft12,\ft12)\,,\nn\\
{\bf m}_3&=&\ft12(0,0,1,0,-\ft12,\ft12,-\ft12,\ft12)\,,\qquad
{\bf m}_4=\ft12(0,0,0,1,-\ft12,-\ft12,\ft12,\ft12)\,,\nn\\
{\bf m}_5&=&\ft12(-1,0,0,0,\ft12,\ft12,\ft12,\ft12)\,,\qquad
{\bf m}_6=\ft12(0,-1,0,0,\ft12,-\ft12,-\ft12,\ft12)\,,\nn\\
{\bf m}_7&=&\ft12(0,0,-1,0,-\ft12,\ft12,-\ft12,\ft12)\,,\qquad
{\bf m}_8=\ft12(0,0,0,-1,-\ft12,-\ft12,\ft12,\ft12)\,.
\eea

\section{The Penrose Inequality for  Time-Symmetric  Data}

In the time-symmetric case, a marginally closed 
outer trapped surface or (MCOTS) 
coincides with what mathematicians call a closed stable minimal 
surface,
that is, one whose second variation is positive.\footnote{For historical 
reasons mathematicians abandon their customary linguistic precision 
and refer to any 
critical  point of the area functional, regardless of the nature of 
its Hessian as a ``minimal
surface.''  The adjective  ``stable'' has no dynamical significance, 
but is taken to mean  that the Hessian is positive definite.}
The apparent horizon  is the outermost MCOTS, and coincides with the outermost  
stable minimal surface \cite{ Hawking:1971vc,Gibbons:1972ym}. 
The area  $A$  of an apparent horizon
is usually taken as a lower bound for the area 
$A_{\rm initial} $ of the intersection
of the event horizon with the initial surface. 
Assuming  cosmic censorship is valid, then
by Hawking's area theorem \cite{Hawking:1971tu,Hawking:1971vc}
this should be no larger
than the area $A_{\rm final}$ of the final black hole. 

   If the final black hole
is non-rotating  and it and carries no electric charges, 
the final state should be a Schwarzschild black hole, whose 
mass $M_{\rm final}$ is given by
\ben
A_{\rm final} = 16 \pi M_{\rm final}^2 \,.
\een 
We also have that the initial ADM mass $M_{\rm initial} $ of the data set 
should satisfy 
\ben
M_{\rm final} \le M_{\rm initial}  \,.
\een 
Thus we expect that 
\ben
A \le A_{\rm initial } \le A_{\rm final } = 16 \pi M_{\rm final}^2 \le 
16 \pi M_{\rm initial} ^2 \,,\qquad \Rightarrow
\qquad A \le  16 \pi M_{\rm initial} ^2 \,. \label{cosmic}
\een
The last inequality in (\ref{cosmic}) is called the 
\emph{Penrose Inequality},  or 
\emph{Cosmic Censorship Inequality} \cite{Penrose:1973um}. 
Moreover, 
one obtains in this way an upper  bound on the efficiency 
\ben
\eta = \frac{M_{\rm initial}-M_{\rm final}} {M_{\rm initial} } 
\le  1- \sqrt{\frac{A}{1 6 \pi M_{\rm initial } ^2} }   
\label{efficiency}
\een
with which the time development of the initial 
data converts rest mass to gravitational radiation. 
Thus although there now exist general proofs 
of the Penrose inequality in the general time-symmetric case
due to Huisken and Ilmanen, based on the inverse curvature flow
proposed for this purpose by Geroch, the value of $\eta$
remains of interest.    

In the discussion above we have assumed that no electric
or magnetic  charges
are carried by the final black hole. If that is not   
so, the bounds are modified. 

  As an illustration of the above idea, we shall now consider the 
example of time-symmetric initial data for an  Einstein-Scalar
system, for which the equations of motion are
\ben
R_{\mu \nu}=2 \p_\mu \phi \p_\nu \phi\,.
\label{ED}
\een
The only time-symmetric initial value constraint is
\ben
{^{(3)} R}= 2 g^{ij}  \p_i \phi \p_j \phi\,.
\een 
As shown by Ortin \cite{Ortin:1995vg},
a set of time-symmetric initial data depending upon one
parameter $\alpha$  is given by
\ben
g_{ij} = C^{2(1-\alpha)}  D^{2(1+\alpha )} \delta_{ij}
 \,,\qquad e^\phi = e^{\phi_0}
\bigl(\frac{C}{D}\bigr)^{\pm\sqrt{1-\alpha^2}}\,,  
\label{Ortindata}
\een
with $C$ and $D$ harmonic and $-1<\alpha <1$. This can also be seen from our
expressions for the Case (1) time-symmetric data for the 
Einstein-Maxwell-Dilaton system in section \ref{Case1sec}, by taking
$a^2=\alpha^2/(1-\alpha^2)$ so that the electric field vanishes.
 (Exchanging $C$ and $D$
sends $\phi\rightarrow -\phi$.) If
\ben
C= 1+ \frac{X}{2 \rho} \,, \qquad
 D= 1+ \frac{Y}{2 \rho} \,,\label{functions}
\een
the initial ADM  mass $M_{\rm initial} $
and initial (non-conserved) scalar charge  
$\Sigma_{\rm initial}$ are given by
\ben
M_{\rm initial} = M =  \half (1-\alpha)X + \half (1+\alpha) Y\,,\qquad 
\Sigma_{\rm initial}= \Sigma = \half \sqrt{1-\alpha^2}(Y-X) \,. \label{mass}
\een

If $X>0$ and $Y>0$,  the solution is regular for $0< \rho <\infty$, 
and near $\rho =0$ we have 
\ben
ds ^2 \approx X^{2(1-\alpha)}\, Y^{2(1+\alpha)}
\frac{1}{16\rho ^4} \,
\Bigl ( d \rho ^2 + \rho ^2 \bigl(d\theta^2 + \sin^2\theta d \varphi^2\bigr)  
\Bigr)\,. 
\een

If we set $R=1/\rho$ we find
\ben
ds ^2 \approx \fft1{16}\, X^{2(1-\alpha)}\, Y^{2(1+\alpha)}\,
 \Bigl ( d R^2 + R^2  \bigl ( d \theta ^2 + \sin ^2 \theta d \varphi^2 
 \bigr) \Bigr ) \,.
\een
Thus $\rho=0$ corresponds to another asymptotically flat
region.
The two asymptotically flat regions are separated by an Einstein-Rosen 
bridge. If $X\ne Y$ the scalar charge is non-vanishing and 
there is, unlike in the Schwarzschild case,
an  asymmetry between the two asymptotic  regions, and
the  values of the scalar fields  at the two infinities
will differ.

There is a unique   totally geodesic two-sphere
at 
\ben
\rho=\rho_+ = \frac{\alpha (X-Y ) }{4 }    +    
\sqrt{  \bigl(\frac{\alpha (Y-X) }{4 } \bigr)^2 +\frac{XY}{4}   } \,,
\label{rhoplus}
\een
located between $\rho=0$ and $\rho=\infty$, at which the area 
\ben
A(\rho) = 4 \pi \rho ^2 \Big(1 + \frac{X}{2 \rho} \Big)^{2(1-\alpha )}\,
 \Big(1 + \frac{Y}{2 \rho} \Big)^{2(1+\alpha )} 
\een  
attains an absolute  minimum.
The  scalar no hair theorem 
\cite{Chase,Bekenstein:1971hc,Hawking:1972qk} implies
that there is no non-singular static black hole with non-constant
scalar field, and the expected final state is a Schwarzschild 
black hole with mass $M_{\rm final} $ and 
vanishing scalar charge  $\Sigma_{\rm final}$.

   We may now consider the Penrose inequality
\be
W\equiv 16\pi M^2 - A(\rho_+) \ge0\,.
\ee
It is helpful to parameterise $X$ and $Y$ in terms of new quantities
$q$ and $s$, such that
\be
\alpha (X-Y)= q\, s \,,\qquad 2 \sqrt{XY} = q \sqrt{1-s^2}\,,
\ee
where 
\be
q\ge 0\,,\qquad -1\le s\le 1\,.
\ee
This gives
\be
\rho_+ = \ft12\alpha\,  q\,  (1-s)\,
\ee
when $\alpha>0$, and $\rho_+ =-\ft12 \alpha \,q \,(1+s)$ 
when $\alpha<0$.
We can focus, without loss of generality, on the case $\alpha>0$, since 
reversing the sign of $\alpha$ is equivalent to switching $X$ and $Y$.
We then have
to prove that 
\be
W\equiv 16 (\sqrt{\alpha^2 (1-s^2) +s^2} -
 \alpha \, s )^2-
\alpha^2(1- s)^2\,  Z_+^{2(1-\alpha)}\, Z_-^{2(1+\alpha)} \ge0\,,
\label{Wdef}
\ee
where 
\be
Z_\pm\equiv 
1+\fft{\sqrt{\alpha^2(1-s^2) + s^2}\pm s }{
                          \alpha(1-s)} \,,
\ee
and where $0<\alpha\le 1$ and $-1\le s\le 1$.  
It is evident that we can take the square root of 
both sides in (\ref{Wdef}), and so we need to show that 
$H\ge 0$ for $0\le\alpha\le 1$ and $-1\le s\le 1$, where
\be
H= 4 \sqrt{\alpha^2 + (1-\alpha^2)\, s^2} - 4 \alpha\, s -
  \alpha\, (1-s)\, Z_+\, Z_-\, R^\alpha\,,\label{Hdef}
\ee
and we have defined
\be
R \equiv \fft{Z_-}{Z_+}\,.
\ee
It is rather straightforward to show analytically that $R$ 
varies monotonically
as a function of $s$, with
\be 
R(-1)= 1+\fft1{\alpha}\,,\qquad R(0)=1\,,\qquad R(1)=0\,.
\ee
$Z_+$ and $Z_-$ are both $\ge 1$.

We have not found an analytic proof that $H$ defined in (\ref{Hdef}) indeed
satisfies $H\ge 0$ but it is evident from numerical analysis that this
inequality is satisfied, and hence that the Penrose inequality holds for the
Einstein-Scalar system with time-symmetric initial data.\footnote{After
circulating an initial version of this paper, David Chow showed us an
analytic proof of the inequality $H\ge 0$.}

\section{Concluding Remarks}

In this paper we have presented time-symmetric initial data results for  the  Einstein-Maxwell-Dilaton theory with a general dilation coupling $a$, which can be  expressed in terms of three harmonic functions. We have also generalised the results to the  Einstein-Maxwell-Dilaton theory with two Maxwell field and presented the result in terms of four harmonic functions. 
The initial data results for the STU model, with four electric fields and three dilation fields  can be expressed in terms of  eight harmonic functions. We also matched the results to know static black hole solutions, and as a by-product, we presented these metrics and all the sources in terms of specific harmonic functions. 

The method can be in general applied to other supergravity models. 
We also showed how for all the theories with electric fields, 
the initial-data could be mapped into initial data for theories with 
scalar fields only.  For example,  the initial data for the 
Einstein-Maxwell-Dilaton theory of section 3  can be  mapped onto the 
initial data problem of an Einstein-Scalar model with two scalars, and 
the initial data for the STU model map onto data for an Einstein-Scalar 
model with seven scalar fields.   We then gave a rather general construction
of time-symmetric initial data for a system of $N$ scalar fields coupled
to gravity.
 
 While the work provides a prerequisite for study of the time evolution of initial data, there are also a number of physical properties one may explore 
without having to evolve the data with the full equations of motion.  
For example,  for initial data  for multi-black hole systems one may,
following \cite{Brill:1963yv}, associate masses and charges with the 
individual black holes and hence one may calculate binding energies. 
We have performed  calculation of the interaction energies for the case of two  multi-centered harmonic functions $C$ and $D$, i.e.
 $\Phi^2=C^\gamma D^\delta$. It turns out that for the general harmonic 
functions, the ADM mass $M_\infty$,  as measured at infinity 
 and  a constant part of the sum of  constituent masses 
$\sum_{a=1}^nM_a$  do not cancel, except 
 for  $\gamma=\delta=1$ which is  the Einstein-Maxwell case.  It is also 
only in this case that the  remaining interaction energies can be cast in 
a form that has a physical interpretation in terms of the  gravitational 
and electric potential energies of the system. This may be due to the fact 
that in other examples the scalar field interactions modify both 
constituent mass contributions as well as the nature of the 
interaction potential energies.  We note, however, that 
the initial  data  for the multi-centered static black hole solutions do 
have a cancellation of the asymptotic and constituent constant mass 
contributions, as expected.  We have also carried out the calculation for
the STU model,  with parallel results.
 
  Another example where one can study physical properties without needing
to evolve the initial data is provided by the Penrose area inequality,
given in (\ref{cosmic}).   As an illustration we
 studied the specific example of a scalar field coupled to gravity, and
we showed by numerical means that indeed the time-symmetric initial data
necessarily gave rise to configurations that satisfy the area inequality.
It would be of interest to generalise this result to other examples 
of time-symmetric initial data, and also to find an analytic proof that
the inequality is obeyed.

We should like to conclude with remarks about results for the initial data, 
not necessarily-time symmetric, that lead to time-dependent solutions.
Specifically,  Nakao, Yamamoto and Maeda \cite{Nakao:1992zc}  pointed
out that initial data for Einstein's vacuum  equations with
a positive cosmological constant $\Lambda=3H^2$ can be constructed by 
starting with the time-symmetric initial data with a non-vanishing 
cosmological constant that is not time-symmetric.
This result  in turn also  leads to time-dependent solutions, 
with a positive cosmological 
constant, such as the Kastor-Traschen multi-centered solutions  
of Einstein-Maxwell-de Sitter gravity \cite{Kastor:1992nn}.
There are by now large classes of   multi-centered extremal cosmological 
black hole solutions known in  fake gauged supergravity theories with a
positive cosmological constant, such as those in the  gauged STU 
model  \cite{Behrndt:2003cx},  as well as non-extremal 
cosmological black hole solutions in  gauged Einstein-Maxwell-Dilaton 
models \cite{Gao:2004tu}.
Further exploration of time-nonsymmetric initial data and the  
time-dependent solutions for  general gauged supergravity theories, 
and especially the intriguing connection to the time-symmetric data of 
the corresponding ungauged supergravity theories, deserves further study.

\vskip 1in
{\noindent\large  \bf Acknowledgments}
\vskip 0.1in
\noindent The work is supported in part by the DOE Grant DOE-EY-76-02-
3071 (MC), the Fay R. and Eugene L. Langberg Endowed Chair (MC), the Slovenian 
Research Agency (ARRS) (MC), and the Simons Foundation Fellowship (MC).
The work of C.N.P.
is supported in part by DOE grant DE-FG02-13ER42020.
 MC would like to  thank Cambridge University for hospitality 
during the course of the work.


\end{document}